\documentstyle[prl,aps,multicol,psfig,epsf]{revtex}
  \renewcommand{\narrowtext}{\begin{multicols}{2} \global\columnwidth20.5pc}
  \renewcommand{\widetext}{\end{multicols} \global\columnwidth42.5pc}
\begin{document}
\title{Disorder and interaction induced pairing in the addition spectra of 
quantum dots}
\author{C.\ M.\ Canali}
\address{$^{1}$Department of Theoretical Physics, Lund University,
  S\"olvegatan 14, 223 62 Lund, Sweden\\}
\date{\today}
\maketitle
\begin{abstract}
We have investigated numerically the electron addition spectra
in quantum dots 
containing a small number ($N\le 10$) of interacting electrons,
in presence of strong disorder.
For a short-range Coulomb repulsion, we find regimes 
in which two successive electrons
enter the dot at very close values of the chemical potential.
In the strongly correlated regime these close additions
or pairing are associated with electrons tunneling into
distinct electron puddles within the dot.
We discuss the 
tunneling rates at pairing and we argue that our results are
related to a phenomenon known as {\it bunching}, recently observed
experimentally.

\end{abstract}
\draft

\narrowtext
In a small metallic island weakly coupled to the environment the
number of electrons is quantized at low temperatures.
Because of the Coulomb repulsion from the electrons already on the island,
it takes a finite energy to add one more electron to the island.
This can be achieved with the help of an
an external gate voltage, coupled capacitatively to the island.
Additions of single electrons occur roughly periodically 
as a function of the gate voltage. This is the essence of Coulomb
blockade (CB), one the most robust facts in 
mesoscopic physics.\hfill\break
\indent
By means of an experimental technique known as 
single-electron capacitance spectroscopy (SECS)\cite{ashoori92,ashooriN},
one can study electron additions from a metallic electrode to a
semiconductor quantum dot.
In these experiments the electrons tunnel into localized states of 
the dot one by one,
starting from the very first electron; each tunneling event is recorded
as a peak in the differential capacitance, measured as a function
of the gate voltage.\hfill\break 
\indent
A few years ago
a SECS experiment\cite{ashoori92,ashooriphysica} showed that 
electrons sometimes entered the dot in pairs rather than individually,
thus violating all the common wisdom that we have on CB.
A more recent and systematic investigation\cite{zhitenev} has shown 
that quite generally, 
in dots containing $N < 200$ electrons, electron additions are
not evenly spaced in gate voltage.
Rather, in some cases, they group in {\it bunches} of up to 6 electrons.
Experimentally the first bunching
occurs already when the number of electrons $N \sim 7$ and with increasing $N$
it evolves from occurring randomly to periodically at about every 5th
electron.
To explain this puzzling phenomenon
two different theoretical models have been proposed\cite{phillips,raikh},
in which two paired electrons enter the dot {\it coherently}.
{\it Classical} simulations\cite{koulakov2,bedanov,levitov} are able 
to reproduce some experimental features observed 
in the presence of a strong magnetic field.
However the mechanism responsible for the pairing effect 
is still unclear.\hfill\break
\indent
The purpose of this work is to carry out a full quantum mechanical
calculation of the addition spectra of dots containing a small number
of particles $N<10$, including their spin. 
Our goal is to investigate the regime where
aperiodic electron pairing first appears. 
We consider the limit of strong disorder
generating localized states that can be occupied by two electrons
of opposite spin\cite{ashooriP}.  
When the Coulomb repulsion is {\it short range},
we find two different regimes
where successive additions can occur {\it sequentially} but
anomalously close to one other. 
The first case takes place for intermediate values of the direct Coulomb
interaction but strong on-site repulsion, which favors the appearance
of a polarized dense droplet with no holes or doubly occupied states. Both
electrons participating in the pairing tunnel into the edges
of the dot but at spatially distinct regions.
The second situation occurs in the strongly correlated regime,
with strong values of the direct 
Coulomb interaction competing with the on-site repulsion and the disorder.
In this case pairing is characterized by the formation of distinct puddles
of electrons; doubly occupied states appear at the center of the dot, 
where one the two electrons tunnels. Specific features of pairing in 
this regime allows us to relate these results to the bunching phenomenon
of Ref.\ \cite{zhitenev}.
\hfill\break
\indent
We model the quantum dot as a finite portion of a square lattice with 
${\cal N}_x \times {\cal N}_x$ sites,
described by the tight-binding Hamiltonian

\begin{eqnarray}
H = &&
\sum_{i,\sigma}(\epsilon_i - eV_g)c^{\dagger}_{i,\sigma}c^{\phantom \dagger}_{i,\sigma}\nonumber\\
&& + t \sum_{\langle i j \rangle,\sigma}
\Big(c^{\dagger}_{i,\sigma}c^{\phantom\dagger}_{j,\sigma}\exp ^{i\,\varphi_{ij}}
+ {\rm H.c.}\Big)\
  +\  H_{\rm int}
\label{hamiltonian}
\end{eqnarray}
where the $\epsilon_i$'s are site energies $\epsilon_i$ randomly distributed 
with uniform probability between $-W$ and $W$; $V_g$ is a gate voltage.
The hopping term, connecting nearest-neighbors sites, 
is proportional
to the matrix element $t$ modified by Peierls phases $\varphi_{ij}$, due
to the possible presence of a magnetic field. 
$\sigma = \uparrow\;,\downarrow$ is a
spin variable.
The interaction part is given by
\begin{equation}
H_{\rm int} = \sum_{i>j\; ; \sigma,\sigma'}{v}({\bf r}_i - {\bf r}_j)
n_{i,\sigma}n_{j,\sigma'} + U\sum_{i}n_{i,\uparrow}n_{i,\downarrow}
\end{equation}
where $n_{i,\sigma}$ is the number operator, ${v}({\bf r}_i - {\bf r}_j)$ 
is the matrix element of the direct part of the Coulomb interaction
and U is the on-site-interaction constant. 
\vfill
\begin{figure}[tbp]
\begin{center}
\hspace{0.2truecm}
\psfig{file=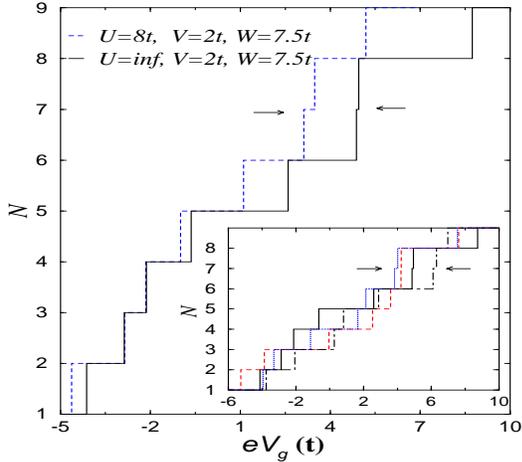,height=2.3in,width=2.3in}
\end{center}
\vspace{0.3truecm}
 \caption{Number of electrons $N$ on a $3\times 4$ dot vs. gate voltage
for one particular disorder realization ($W = 7.5t$).
and nearest-neighbor Coulomb interaction ($V = 2t$).
The on-site-interaction strength is $U=8t$ (dashed line) and $U=\infty$
(solid line) respectively.
The arrows point to the {\it pairing} between the additions 
$6\Rightarrow 7$ and $7 \Rightarrow 8$. 
Inset: $N$ vs $V_g$
for four disorder realizations of polarized electrons ($U=\infty$),
displaying pairing between the $N=7$ and $N=8$ states.}
\label{fig1}
\end{figure}
\noindent
Experimentally the Coulomb
interaction is screened by the external gate.
For simplicity we will model the interaction
as a nearest-neighbor repulsion, 
${v}({\bf r}_i - {\bf r}_j) = V\delta_{{\bf r}_j, {\bf r}_i+{\bf \delta}}
= {e^2 \over a}\delta_{{\bf r}_j, {\bf r}_i+{\bf \delta}}$,
where $a$ is the lattice constant\cite{note1}. 
The important parameter describing the strength of the average 
Coulomb interaction
relative to the Fermi energy is $r_s = 1/\sqrt{\pi n} a_B$, where $n$ is the
electronic density and $a_B$ is the Bohr radius. For our lattice model
$r_s = {2\over \sqrt{\pi\,\nu}}\,(V/4t)$, where
$\nu= \langle\hat  N\rangle/[({\cal N}_x -1)({\cal N}_y -1)]$,
and $\langle \hat N\rangle$  is the average number of electrons on the dot.
 
By using Lanczos techniques,
we have diagonalized the Hamiltonian Eq.\ (\ref{hamiltonian})
on a $3\times 4$ dot, within the Hilbert subspaces
corresponding to a fixed total number of electrons between $1$ and $10$.
The ground-state energy and wave function,
the total number of electrons 
$N= \langle \hat N\rangle$ and
site occupation $\langle n_{i,\sigma}\rangle$, can be obtained for 
different values of the parameters entering Eq.\ (\ref{hamiltonian}).
$N$ is controlled by the voltage $V_g$.
At a specific value $V_g^{N+1}$
given by
\begin{equation}
e\, V_g^{N+1} = E_0^{N+1} - E_0^N \equiv \mu(N+1)
\label{chemp}
\end{equation}
the number of electrons on the dot jumps from $N$ to $N+1$.
Here $E_0^N$ is the ground-state energy for $N$ particles at $V_g =0$;
$\mu(N)$ is the chemical potential, which
is the quantity measured experimentally\hfill\break
\indent
In Fig.\ \ref{fig1} we plot the dot occupation $N$ as a function
$V_g$ for one random configuration in the regime of strong
disorder ($W = 7.5t$) and intermediate values of Coulomb interaction $V=2t$,
corresponding to $r_s \approx 0.5$. For onsite
repulsion $U=8t$ (dashed line),
the two additions $6\Rightarrow 7$ and $7\Rightarrow 8$ occur at close
gate voltages; in contrast, all other electrons enter the dot at
well-spaced voltages. 
\vfill
\begin{figure}[tbp]
\begin{center}
\hspace{0.0truecm}
\psfig{file=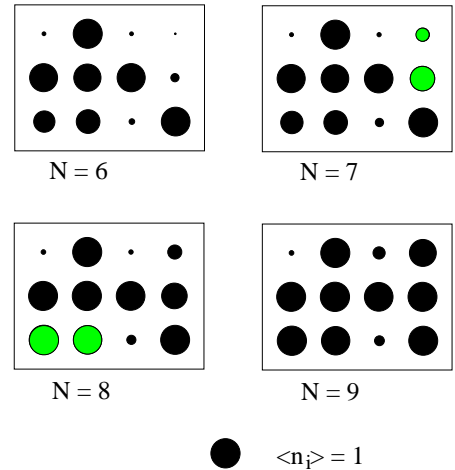,height=2.5in,width=2.3in}
\end{center}
\caption{Ground-state site occupation $<n_i>$
for different values of $N$, for the dot of Fig.\ \ref{fig1} (solid line).
The area of the circles is proportional $<n_i>$.
Pairing occurs between the $N=7$ and $N=8$ states.
The grey circles indicate the
sites where the largest portion of the incoming electron,
participating in the pairing, is distributed.}
 \label{fig2}
\end{figure}
\noindent
Moreover, the tendency of electrons 7 and 8 to pairing
is strongly enhanced by increasing U. This is clearly shown 
by the solid line of 
Fig.\ \ref{fig1} where $U=\infty$, corresponding to 
{\it spin-polarized} electrons.
In this case the two pairing
particles still enter the dot sequentially but almost at the same gate voltage.
Since this result refers to one particular disorder realization, one might
think that it represent an extremely rare episode. Remarkably however,
this is not the case, as the inset of Fig.\ \ref{fig1} shows: out of 8
random realization that we have tried, half of them displays paring
between the additions $6\Rightarrow 7$ and $6\Rightarrow 8$, 
when the system is polarized,
i.e. $U=\infty$.\hfill\break
\indent
In order to understand how pairing takes place,
we look at how the site occupation $<n_i>$ changes
when the two participating particles tunnel into the dot.
Fig.\ \ref{fig2} displays $<n_i>$ for the $3\times 4$ sites of the
dot, corresponding to the polarized case of Fig.\ \ref{fig1}. Because
$U=\infty$, no double occupancy is allowed and the electrons tend to form
one rather compact puddle (see the state $N=6$ in Fig.\ \ref{fig2}). 
This is reminiscent of
the effect of the exchange interaction studied within the Hartree-Fock
approximation\cite{mdd} for a dot in a strong magnetic field: exchange
generates a local attraction between the electrons,
causing the formation of a dense droplet.
The 7th electron, the first involved in the pair,
tunnels mainly into two almost empty sites at the right edge of 
this compact puddle, as shown by the grey circles of the N=7 state.
The 8th electron, the second involved in the pairing, fills up two 
already partially filled sites on the bottom edge of the dot.
Thus, the pairing of Fig.\ \ref{fig1} is associated with
electron additions into spatially distinct regions of a compact 
electron puddle. 
Because the Coulomb
interaction is short range, the energy costs of these two additions
can be almost equal.\hfill\break
\indent
\vfill
\eject
\begin{figure}
\begin{center}
\hspace{0.2truecm}
\psfig{file=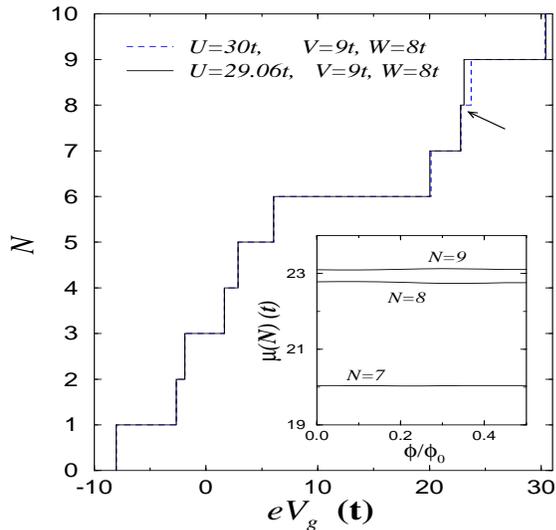,height=2.3in,width=2.3in}
\end{center}
\vspace{0.5truecm}
\caption{Number of electrons $N$ on a $3\times 4$ dot vs. gate voltage
for one particular disorder realization ($W = 8t$)
and nearest-neighbor Coulomb interaction ($V = 9t$).
The two lines correspond to an
on-site-interaction strength $U=30t$ (dashed line) and $U=29.06t$
(solid line) respectively.
The arrows points to the additions $7\Rightarrow 8$ and $8\Rightarrow 9$, 
where pairing occurs.
The two lines differ perceptibly only at pairing.
For the system relative to the solid line,
the inset displays  $\mu(N)$ vs the magnetic flux $\phi$ for $N=7,8,9$.}
 \label{fig3}
\end{figure}

\begin{figure}
\begin{center}
\hspace{0.2truecm}
\psfig{file=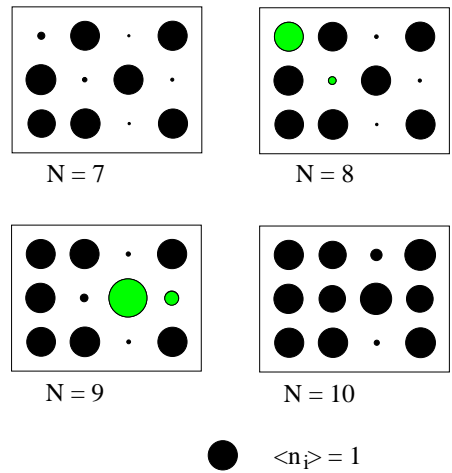,height=2.5in,width=2.3in}
\end{center}
\caption{Ground-state site occupation $\langle n_i \rangle$ of
the $3\times4$ dot
of Fig.\ \ref{fig3} (dashed line, $U=30t$).
Pairing occurs between the states $N=8$ and $N=9$.
Most of the second electron participating in the pair ($N=9$)
tunnels into a central site, which becomes
partially doubly occupied, $\langle n \rangle = 1.70$.}
\label{fig4}
\end{figure}

The results shown above hold for intermediate values of the
Coulomb interaction ($V\approx 2t \Leftrightarrow r_s \le 1$),
much smaller than the strength of disorder
and on-site repulsion. 
Experimentally however, pairing occurs at larger
values Coulomb repulsion ($r_s \ge 2$).
If we increase the strength of the direct Coulomb interaction relative
to the on-site repulsion and the disorder,
we reach the limit where the dot finds more advantageous to generate
local singlets of doubly occupied localized states. Instead of grouping
into one compact puddle, the electrons now can form distint puddles.
Electron pairing can also take place in this 
strongly correlated regime.\hfill\break
\indent
An example is shown in Fig.\ \ref{fig3}, where we plot $N$ vs $V_g$
for a strength of the direct Coulomb repulsion
$V =9t$ ($\Leftrightarrow r_s= 2$), close to the disorder strength, $W=8t$. 
The two curves
corresponds to two values of the on-site repulsion, $U =30t$ and $U=29.06t$. 
We focus on the additions $7\Rightarrow 8$ and
$8\Rightarrow 9$, where pairing takes place\cite{note2}. 
Note that the two curves differ
only in this region, indicating that the on-site repulsion $U$ plays 
a crucial role. We can again understand how this comes about by looking
at the site occupation $\langle n_i \rangle$ for the states involved 
in the pairing. This is shown in Fig.\ \ref{fig4} for $U=30t$.
The 8th electron, the first involved in the pair, enters the dot at the
top left corner of the dot. The 9th electron jumps mainly into one
of the central sites, which becomes partially {\it doubly occupied}.
The dot shows a tendency to create two separate puddles of electrons,
one composed of the 5 occupied sites
along the left edge,
and the second made up  by the
3 sites of the right edge plus
the doubly occupied site.
Each electron involved in the pair tunnels into one of 
\vfill
\noindent
these spatially distinct
regions. 
Finally, when the 10th electron enters the dot, the gap between
the two regions is filled and the dot is occupied more uniformly.
This example supports the suggestion, borne out of the experiment, that
pairing is associated with electron localization in distinct
puddles of the dot\cite{zhitenev}. The merging of the two puddles
upon increasing $N$, corresponds to a sort of 
localization-delocalization transition. \hfill\break
\indent
If $U$ is slightly smaller, e.g. $U=28t$, the roles of the 8th and
the 9th electrons are interchanged: the first electron of the pairing tunnels
into the central site, creating the spin-singlet; next, the second electron
occupies the top edge site. By tuning $U$ between $30t$ and $28t$,
we have found that the intermediate value $U=29.06t$ 
gives rise to the closest additions $7 \Rightarrow 8$, $8\Rightarrow 9$.
These two events still take place sequentially, (see Fig.\ \ref{fig3}).
For this value of $U$, an analysis of $\langle n_i \rangle$ reveals that, 
in each tunneling event participating in the pairing,
half electron goes into
the central site and the other half goes into the top-left corner site.
We have checked that other disorder realizations 
display a similar behavior.\hfill\break
\indent
The inset of Fig.\ \ref{fig3} displays the magnetic flux $\phi$ dependence
of the chemical potential, Eq.\ \ref{chemp}, for the system corresponding
to the ``best pairing'' ($U=29.06t$). 
For such strong values of the interaction and the disorder, 
the traces of the addition spectrum
are only very weakly dependent on $\phi$. This implies that
the pairing
states $N=8$ and $N=9$ remain close to each other for the entire range
of $\phi$.
Experimentally, in bunchings occurring at low $N<10$,
the traces also stick together for the entire range of 
$\phi$,
but at the same time they rise nearly linearly, showing that
the field dependence of the localized states approaches the
one of the lowest Landau level.\hfill\break
\indent
So far we have discussed the occurrence of pairing only
\vfill
\eject
\begin{figure}
\begin{center}
\hspace{0.2truecm}
\psfig{file=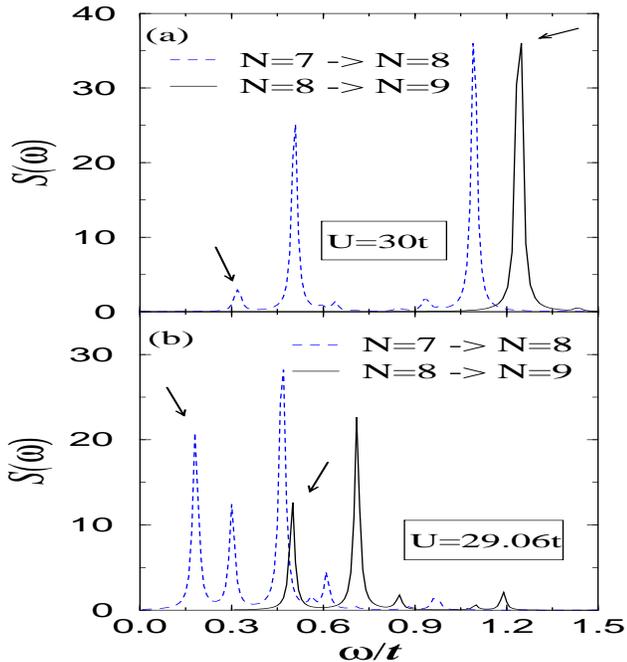,height=3.0in,width=2.3in}
\end{center}
\vskip 0.5truecm
\caption{The addition spectral function for the two particles forming
a pair as in Fig.\ \ref{fig3}. The dashed and solid lines refer
to the first and second
particle of the pair respectively. The peaks indicated by the arrows are
proportional to the tunneling rates for the $N \to N+1$ process.
(a) and (b) refer to dashed and solid line of Fig.\ \ref{fig3} respectively.}
  \label{fig5}
\end{figure}
\noindent
from the point of view of energy balance. We conclude our
analysis by discussing the rates at which the two paired electrons
tunnel into the dot in the strongly correlated regime $r_s \ge 2$.
To this purpose we have computed the addition
spectral function, defined by
\begin{equation}
S(\omega) = \sum_n \Big |
\langle \Phi^{N+1}_n |\sum_{i,\sigma} c_{i,\sigma}^{\dagger}| \Phi^{N}_0\rangle
\Big |^2
\delta \big[\omega - (E^{N+1}_n - E^N_0)\big ]
\label{spectralf}
\end{equation}
where $|\Phi^{N+1}_n\rangle$ are the eigenstates of the
$(N+1)$-system corresponding to energy $E^{N+1}_n$. The operator
$\sum_{i,\sigma} c_{i,\sigma}^{\dagger}$ in Eq.\ \ref{spectralf}
implies that electrons tunnel coherently into all the  sites of the dot 
with equal probability. 
The term in Eq.\ \ref{spectralf}
involving
$\langle \Phi^{N+1}_0 |\sum_{i,\sigma}c_{i,\sigma}^{\dagger}|\Phi^{N}_0\rangle$
is proportional to the rate of tunneling from the ground state
of the $N$- to the $(N+1)$-particle system.\hfill\break
\indent
Fig.\ \ref{fig5} displays the spectral function for the addition of
the two pairing electrons, corresponding to the situation of Fig.\ \ref{fig3}.
The peaks pointed by the arrows are proportional to the tunneling rates.
In Fig.\ \ref{fig5}(a) we can see that for $U=30t$
the tunneling of the first
particle forming the pair is {\it much smaller} than the 
second. As discussed above, in this case the first particle is mainly
localized into one corner site of the dot (see Fig.\ \ref{fig4}).
The presence of fast tunneling in one state of the pair
is also seen experimentally,
albeit starting at strong magnetic field and at higher $N$.
At low $N$ the experiment shows that the tunneling rates
of {\it both} particles are large for all magnetic fields.
As shown in Fig.\ \ref{fig5}(b) the two tunneling rates are
{\it both large} when $U$ takes on the value that ensures the {\it closest
additions}. This can be understood if we remember that in this case
the electronic densities of the pairing particles are distributed
among several sites of the dot. Thus, even if the system 
is strongly correlated,
the tunneling rates for the pairing states are not negligible,
in agreement with the experiment. In contrast, the two models 
in Refs.\ \cite{phillips} and \cite{raikh} predict a dramatic suppressions 
of the tunneling rate at pairing.
\hfill\break
\indent
In conclusion, our numerical simulations show evidence of 
pairing in the addition spectra of disordered quantum dots
with strong on-site repulsion.
Specific features of the pairing observed in the strongly correlated
regime $r_s\ge 2$ suggest that our results
may be related to the analogous phenomenon seen experimentally
in the regime of a small number of electrons.
\hfill\break
\indent
I thank R.\ Ashoori for explaining some experimental aspects,
W.\ Stephan for precious help with the Lanczos code and
Arne Brataas and P.\ Johansson for discussions.

\widetext
\end{document}